\documentclass[twocolumn,showpacs,showkeys,prb]{revtex4}
\usepackage{amsmath,amsfonts}
\newcommand{\alp}{{\alpha}}
\newcommand{\om}{{\omega}}

\newcommand{\bea}{\begin{eqnarray}}
\newcommand{\beq}{\begin{equation}}
\newcommand{\eea}{\end{eqnarray}}
\newcommand{\eeq}{\end{equation}}

\begin{document}
\title
{Non-Hermitian Hamiltonians, Metric, Other Observables and Physical Implications}
\date{\today}
\author{H.B.~Geyer$^{1,2}$, W.D.~Heiss$^{1,2}$ and F.G.~Scholtz$^{1}$}

\affiliation{$^1$Institute of Theoretical Physics, University of Stellenbosch, 7602 Matieland, South Africa\\ and\\
$^2$National Institute for Theoretical Physics (NITheP), Stellenbosch Institute for Advanced Study, 
7600 Stellenbosch, South Africa  }

\begin{abstract}
The metric associated with a quasi-Hermitian Hamiltonian and its
physical implications are scrutinised. Consequences of the 
non-uniqueness such as the question of the
probability interpretation and the possible and forbidden choices of
additional observables are investigated and exemplified by specific
illustrative examples.  In particular it is argued that the improper identification of 
observables lies at the origin of the claimed violation of the brachistchrone transition 
time between orthogonal states. The need for further physical input to remove ambiguities is pointed out.
\end{abstract}
\maketitle

\section{Introduction}
\label{intro}
Quantum mechanics is formulated within the mathematical
framework of a Hilbert space with linear operators acting on the
elements of a Hilbert space - the physical state vectors. Traditionally,
the linear operators considered are Hermitian operators (in the $L^2$ or Dirac 
sense of transposition and complex conjugation), if they
describe quantities called (physical) observables\footnote{Strictly speaking they 
should be self-adjoint operators. Hermiticity 
only requires $(\phi,A\psi)=(A\phi,\psi)$, while self-adjointness
requires that the domains of $A$ and $A^\dagger$ also coincide, 
i.e., self-adjointness is a strict operator equality $A=A^\dagger$}. 
In fact, the mathematical
structure of Hermitian operators is well understood in that they
guarantee a real spectrum. Of equal importance is the complete and
orthogonal eigensystem in that it allows an unambiguous probability
interpretation of an arbitrary state vector: the expansion coefficients
of such vector in terms of the eigensystem are the probability
amplitudes for finding the associated eigenvalues in a measurement.

Already in the early days of quantum mechanics the possible occurrence
of non-Hermitian operators has been encountered. It was realised that an
open system can be described by an effective non-Hermitian Hamiltonian
where dissipation - complex eigenvalues with negative imaginary part -
is brought about by a non-Hermitian term in the Hamiltonian. But even
Hamiltonians that appear to be Hermitian naively can turn out to be
non-Hermitian when boundary conditions are properly taken into
account. Here we mention the Dirac Hamiltonian with the Coulomb potential
$-Z\alpha /r$ which fails to have a unique self-adjoint extension\cite{Rafels} if
$Z\alpha >1$ , or the cranked anisotropic oscillator
which has singularities associated with non-Hermitian
properties, if the rotational frequency is equal to one of the
oscillator frequencies \cite{heinaz}. In these and similar cases
attempts have always been made to relate the onset of the change in
mathematical structure with a physical effect: the ``sparkling of the
vacuum'' in the case of the Dirac equation\cite{Greiner} for $Z\alpha =1$,
or the onset of an instability for the cranking model\cite{heinaz} if
$\Omega_{rot}=\omega_{ho}$. As we see below, a new class of
non-Hermitian Hamiltonians has given rise to different speculations,
the critical review of which forms part of the present paper.

It has been established that Hamiltonians that are symmetric
under $\cal{PT}$ - the product of parity and time reversal  
\cite{bender,benderreview} - can have a real spectrum \cite{dorey,shin} even though such
operators need not be self-adjoint. In a host of papers devoted to the 
subject (see e.g.\ \cite{JPASpecialIssue}, as well as this present Special Issue, for recent developments) 
the term {\em pseudo-Hermitian} operator was coined \cite{mostaf}, particularly related 
to issues concerning the reality of spectra. A
property denoted as {\em quasi-Hermitian} was investigated in earlier work
\cite{sgh} where a complete and fully consistent quantum mechanical framework, based on 
non-Hermitian Hamiltonians treated in conjunction with other observables, had been established.
(In essence this is pseudo-Hermiticity restricted to positive definite norms of states, as required by the probability interpretation which is central to quantum mechanics.)

The present paper revisits the relevant mathematical properties of
quasi-Hermitian operators and critically discusses their possible
physical implications. Particular emphasis is placed upon the metric
associated with non-Hermitian operators and the related characterisation of
the Hilbert space. The metric is introduced in the following section, with an emphasis
on the notion of physical observables and their consistent introduction into a quantum mechanical framework
built on a non-Hermitian Hamiltonian. In Sect. \ref{prob} we emphasise that the probability interpretation of quantum mechanics breaks down if care is not exercised in identifying physical observables within the quasi-Hermitian context. We also use the proper prescription for calculating transition probabilities with quasi-Hermitian observables to analyse when and why certain transition probabilities will not require a unique construction of the metric, while others do. In Sect. \ref{examples}  we illustrate these points by way of a few simple examples. This includes physical implications relating to the role of observables and the speculation about the violation of the lower limit for transition times between orthogonal states of (Hermitian) observables (the ``non-Hermitian brachistochrone problem"). A summary concludes the paper.

\section{The metric and observables}
\label{sectmetric}

Even though the properties listed below have been stated and analysed in the
literature \cite{sgh,mostaf} we here give a short recapitulation, with somewhat 
different emphasis, to make the paper self-contained. 

While unbounded non-Hermitian operators constitute a
difficult mathematical problem in general, there is the much simpler
class that is well understood: the {\em quasi-Hermitian} operators.
A non-Hermitian operator $H\ne H^{\dagger}$ is called quasi-Hermitian
if there exists a bounded Hermitian positive-definite operator 
$\Theta $ that ensures the relation
\beq
\Theta H=H^{\dagger }\Theta.  \label{qh}
\eeq
The relation implies that $H$ is similarly equivalent to a Hermitian
operator, in fact the operator
\beq
h_S=SHS^{-1} \label{sim}
\eeq
is Hermitian when $S$ is the positive root of $\Theta $. The
operator $\Theta $ may be viewed as a metric characterising a
different Hilbert space by defining the scalar product 
\beq
\langle \cdot |\cdot \rangle _{\Theta }:=\langle \cdot|\Theta \cdot
\rangle \label{scal}
\eeq
where $\langle \cdot|\cdot \rangle $ is the usual scalar
product, employing the $L^2$-metric being the identity.
It is easily verified that (\ref{qh}) guarantees that the non-Hermitian $H$ 
(with respect to the inner product $\langle \cdot|\cdot \rangle $)
is Hermitian with respect to $\langle \cdot|\cdot \rangle _{\Theta }$.
Moreover, in view of (\ref{sim}) the spectrum of $H$ is real. Note,
however, that the eigenvectors $|n\rangle $ obeying
$H|n\rangle =E_n|n\rangle $ are generally not orthogonal 
with respect to $\langle \cdot|\cdot \rangle $ . There is, however,
the bi-orthogonal system using the left hand eigenvectors
$\langle \tilde n|H=\langle \tilde n|E_n$ with
$\langle \tilde n|m\rangle =N_n\delta _{nm}$ which can be written as
$\langle n|\Theta m\rangle =N_n\delta _{nm}$. In view of the
discussion below we stress here the obvious: vectors being orthogonal with
respect to a $L^2$-metric are not orthogonal when using
the $\Theta $-metric and {\it vice versa}; the structure of a Hilbert space, that is the
norm and orthogonality of its elements as well as the concept of
Hermiticity is given by the definition of a scalar product.

The analysis of a quantum system described by a non-Hermitian Hamiltonian $H$
usually confronts the construction and role of the metric $\Theta$ by first focussing on the condition
(\ref{qh}) for $H$. However, it should be emphasised that (\ref{qh}) does not uniquely determine 
$\Theta$. It is simple to see \cite{sgh} that if $\Theta$ is an acceptable metric determined by condition (1), 
so is $\tilde\Theta=V\Theta$ with $V$ any operator that commutes with $H$. $\Theta$ is 
only uniquely determined when condition (1) holds for an  {\it irreducible set} of operators $A_i$, {\it all} 
of which satisfy $\Theta A_i=A_i^{\dagger }\Theta $ for the {\it same} $\Theta$. This also determines 
those operators which may serve as physical observables (see \cite{sgh}). 

This observation is not only important to identify observables, but by implication also for the characterisation 
or labelling of states in terms of quantum numbers associated with observables, as well as for the proper calculation of transition matrix elements, as we proceed to illustrate in the following two sections.

To conclude this section we point out that the non-uniqueness of the metric as determined by a non-Hermitian (quasi-Hermitian) Hamiltonian {\em only}, leaves open the question of the definition (or construction) of other observables which may act consistently with it. Naturally, once a particular metric $\Theta$ conforming to (\ref{qh}) has been chosen (rather than being {\em dictated} by an {\em a priori} consistent choice of a set of observables), then the observables can be obtained by {\em inverse} similarity transformation from standard observables, say $a_i$ (such as $x$ and $p$). The operators so obtained, 
\beq
\label{hermobs}
A_i = S^{-1}a_iS
\eeq
are then by construction quasi-Hermitian w.r.t. the metric $\Theta$ which defines $S$, and accordingly qualify as observables.

\section{Probability interpretation and matrix elements} 
\label{prob}
As indicated in the previous section, orthogonality of the eigenstates of a quasi-Hermitian
operator is defined (and guaranteed) with respect to the inner product 
$\langle \cdot |\cdot \rangle _{\Theta }:=\langle \cdot|\Theta \cdot\rangle$. Moreover, 
a given or chosen $\Theta$, compatible with (\ref{qh}), determines an irreducible set 
of observables $A_i$, and {\em vice versa} \cite{sgh}.

It is almost pedantic, but nevertheless important in view of some recent discussions in the literature \cite{brachi1,brachi2,brachi3}, to emphasise the following. A cornerstone of quantum mechanics, the probability interpretation for measurement results of a given observable, rests on the orthogonality of its eigenstates. It is important to realise, therefore, that if an operator, typically an observable (say $B$) Hermitian with respect to the standard inner product
$\langle \cdot|\cdot \rangle $, is {\em not} from the irreducible set $A_i$ determined by $\Theta$, then any standard calculation and interpretation of matrix elements involving eigenstates of $B$, would be meaningless within the context of the quantum mechanical system described by the non-Hermitian Hamiltonian $H$ and observables $A_i$. This remains so even when $B$ has a real spectrum (which of course does not depend on the Hilbert space inner product) and eigenstates which are orthogonal with respect to $\langle \cdot|\cdot \rangle $. The reason is simply that a consistent calculation of matrix elements (or probabilities) within the $\Theta$ context requires the use of the inner product $\langle \cdot |\cdot \rangle _{\Theta }:=\langle \cdot|\Theta \cdot\rangle$. Hence, absurd conclusions result upon using $B$ as though it were a proper observable; e.g.\ denoting the eigenvalue problem for $B$ as $B|\beta_m\rangle = b_m|\beta_m\rangle$, the probability of finding an eigenvalue $b_m$ upon measurement of $B$ in {\em eigenstate} $|\beta_n\rangle$ ($m\neq n$) is $\langle \beta_m |\Theta|\beta_n\rangle$ which generally $\neq 0$. For a recent critical analysis of this point see \cite{zurek}.  (Similar arguments have recently also been put forward by Mostafazadeh \cite{ali2007prl}.)

Even when calculations are consistently performed, i.e.\ with due recognition of the metric operative within a given context, one may investigate the (in-)dependence of physical results on the choice of a metric determined by a non-Hermitian Hamiltonian as in (\ref{qh}), in particular from the point of view that $\Theta$ is not uniquely determined by (\ref{qh}).

Let us accordingly consider an arbitrary member $A$ from the set of observables which has determined $\Theta$ uniquely. (It is worthwhile to stress that, at least in principle, either a given $\Theta$ is chosen from among those that satisfy condition (1), which in turn uniquely determines which operators qualify as observables, or alternatively, the (non-Hermitian) Hamiltonian is supplemented by operators pre-chosen to be observables (and which form an irreducible set), which in turn pins down $\Theta$  -- see \cite{sgh} for a full discussion). Here we simply assume that $\Theta$ has been fixed and that $A$ is an admissible observable. The prescription for calculating the transition probability $P_{nm}(A)$ for the operator $A$ between eigenstates 
$|m\rangle $ and $|n\rangle $ of the Hamiltonian $H$ is unambiguous and yields the following:

\bea
P_{nm}(A) &=&\frac{|\langle n|\Theta A |m\rangle |^2}{\langle n|\Theta|n\rangle \langle m|\Theta|m\rangle }\nonumber\\
&=& \frac{\langle n|\Theta A|m\rangle^* \langle n|\Theta A|m\rangle }{\langle n|\Theta|n\rangle \langle m|\Theta|m\rangle }\nonumber\\
&=& \frac{\langle m|\Theta A|n\rangle \langle n|\Theta A|m\rangle }{\langle n|\Theta|n\rangle \langle m|\Theta|m\rangle } \; .
\eea

Here $\langle n|$ is the bra or functional associated with $|n\rangle$ through the original $L^2$
inner product. In the second line we have used the fact that $\Theta$ is Hermitian (in the original sense) and that $A$ satisfies the quasi-Hermiticity relation $\Theta A=A^{\dagger }\Theta $.

Let us assume for the moment that the metric had not yet been uniquely fixed, e.g.\ when a full set of irreducible observables has not yet been specified, but with both $H$ and $A$ already included. In this case the calculation above may be repeated for another admissible metric $\tilde\Theta$, to give 

\bea
P_{nm}(A)&=& \frac{\langle m|\tilde\Theta A|n\rangle \langle n|\tilde\Theta A|m\rangle }{\langle n|\tilde\Theta|n\rangle \langle m|\tilde\Theta|m\rangle }\nonumber\\
&=& \frac{\langle m|V\Theta A|n\rangle \langle n|V\Theta A|m\rangle }{\langle n|V\Theta|n\rangle \langle m|V\Theta|m\rangle }\nonumber\\
&=& \frac{\langle m|\Theta A|n\rangle \langle n|\Theta A|m\rangle }{\langle n|\Theta|n\rangle \langle m|\Theta|m\rangle },
\eea
the same result as before! Here we have used the fact that $V=\tilde\Theta\Theta^{-1}$ commutes with $H$, whence the states $|n\rangle$ are simultaneous eigenstates of $H$ and $V$ (if there are no degeneracies in the spectrum), yielding a cancellation of eigenvalues above and resulting in a transition probability which is {\it independent} of the metric!

However, it should immediately be cautioned that this is not a general result (see also \cite{sgh}). It depends on (a) the absence of degeneracies in the spectrum, and more importantly (b) on the fact used above that the transition probability has been calculated with respect to eigenstates of $H$ for an operator $A$ that shares the property of quasi-Hermiticity with respect to a common metric (albeit not a unique one). Furthermore, it is clear that for general states (not eigenstates of $H$ or another operator which shares in the property of quasi-Hermiticity) interference terms will appear which explicitly depend on the metric. If the metric had been chosen with respect to the Hamiltonian only, physical results will in general depend on the particular choice. 

Writing the results (4) and (5) in the form
\bea
P_{nm}(A) &=& \frac{\langle m|\Theta A|n\rangle \langle n|\Theta A|m\rangle }{\langle n|\Theta|n\rangle \langle m|\Theta|m\rangle }\nonumber \\
&=& \frac{\langle \tilde m| A|n\rangle \langle \tilde n| A|m\rangle }{\langle \tilde n|n \rangle \langle \tilde m|m\rangle }
\eea
shows that a transition probability between eigenstates of $H$ (or of any one of the observables) may be obtained by explicitly introducing both right and left eigenstates; choosing a particular bi-orthonormal normalisation 
$\langle \tilde n|m\rangle = \langle n|\Theta m\rangle = \langle n|\Theta| m\rangle = N_n\delta _{nm}$ is equivalent to introducing a particular (and corresponding) metric or inner product. 

The above point seems to have been overlooked from time to time in the literature, when the impression is given that the metric simply dictates a simultaneous transformation of operators and states, with no physical implications. It is also closely related to another simple, but important observation: when matrix elements or transition probabilities are being discussed or explicitly calculated within a quasi-Hermitian framework, the characterisation (or labelling) of such states can only be meaningfully carried out with reference to a set of operators (which should provide relevant quantum numbers) which are all observables and hence quasi-Hermitian with respect to the same metric. We again illustrate this point in section 5.

\section{Illustrative examples}
\label{examples}

\subsection{Non-uniqueness of the metric}

The existence of a metric for a given non-Hermitian yet
quasi-Hermitian Hamiltonian has been demonstrated in numerous 
examples \cite{sgh,gey,jones,sgplb,sg,ali,musumbu}, and ways of constructing it have been
given \cite{sg,sgplb,jones,ali,fringfaria,alimetric,musumbu}. An important aspect that has already been
emphasised in \cite{sgh} and in the two previous sections seems to have been of lesser concern for some
of the authors quoted. It is the fact that there are many possible 
metric operators that ensure (\ref{qh}), 
usually a whole continuum. In other words, there are many ways of
finding a Hermitian Hamiltonian $h_S$, each associated with a
different metric, i.e.~each invoking a different Hilbert space. 
The physical implications are discussed in the following section. Here we focus upon some formal
issues.

A simple example \cite{musumbu} may demonstrate the essential
points. Consider the toy Hamiltonian
\beq
H=\om  (a^{\dagger }a +\frac{1}{2}) +  \alp a^2 + \beta a^{\dagger 2}  \label{ham}
\eeq
with $a$ the usual boson operators. For $\alpha \ne \beta $ the
Hamiltonian is manifestly non-Hermitian. The spectrum is given by
$E_n=(n+1/2)\Omega $ with $\Omega  =\sqrt{\om ^2-4 \alp \beta }$ which is real for 
$\om ^2>4\alp \beta $. For real $\Omega $ a continuum of choices can
be found for the metric and can be
parametrised by the real parameter $z\in [-1,1]$ yielding
\beq
h_{S(z)}=\Omega (b_z^{\dagger}b_z^{}+\frac{1}{2})\equiv
\frac{1}{2}(\mu(z)p^2+\nu(z)x^2).  \label{herm}
\eeq
with $\mu(z) \nu(z)=\Omega^2$ identically in $z$.
The metric by which this is achieved is given by \cite{musumbu}
\beq
\Theta(z)
 =\bigg(\frac{\alp +\beta-\om z+(\alp -\beta)\sqrt{1-z^2}}{\alp
    +\beta-\om z-(\alp -\beta)\sqrt{1-z^2}}
\bigg)^{\frac{ p^2(1-z)+\om ^2 x^2(1+z )}{4\om \sqrt{1-z^2}}}.
\label{simz}
\eeq
up to a constant.
Being faced with a continuous variety of Hilbert spaces ${\cal H} _z$ the
obvious question is now for Hermitian operators other than $H$
(recall that $H$ is Hermitian in ${\cal H} _z$ for any $z$). The
answer is readily given by (\ref{simz}) and will, of course, depend on
$z$. In fact, seeking the operator $O$ such that
$$\Theta(z)O=O^{\dagger }\Theta(z) $$
there is only the combination $O(z)=p^2(1-z)+\om ^2 x^2(1+z)$
fulfilling this requirement. Note that neither the momentum nor the position
operator remains Hermitian except if $z=-1$ or $z=1$, respectively;
for $z=0$ we identify the number operator 
$a^{\dagger }a\sim p^2+\om ^2 x^2$ as being Hermitian.

This example nicely demonstrates the point made in \cite{sgh}: the
metric can be defined uniquely if operators in addition to $H$ are
required to be quasi-Hermitian. In general the additional operators have
to form an irreducible set together with $H$, in the example
above there is just one more, namely $O(z)$.

While the expressions (\ref{herm}) and (\ref{simz}) are valid only if
$\Omega $ is real, there are further restrictions from the requirement
that $\Theta $ be bounded and strictly positive. Yet the numerator or
denominator of (\ref{simz}) vanishes for
\beq
z_{\pm}=\frac{(\alp +\beta)\om  \pm(\alp -\beta)\Omega}{\om 
^2+(\alp -\beta )^2},  \label{zcrit}
\eeq
respectively. For $z$ between $z_-$ and $z_+$ the metric is ill
defined and $h_S$ is not the Hermitian equivalent of $H$. These
singularities are spurious, i.e.~they are not intrinsic to the problem
and can be removed \cite{sg}. In the example given the substitution of $z$ by
$-z$ achieves this goal but at the expense of trading in spurious singularities
elsewhere. The conclusion is that a metric that is globally free of
singularities cannot be found \cite{sg}.

There are other more serious types of singularities as they are
intrinsic to the problem and cannot be removed.
They always render the metric undefined. These are the exceptional
points (EP) \cite{kato,heiss,heissjpa} occurring - in
the example - for
$\Omega =0$, the point where levels coalesce and 
the spectrum becomes non-real. We return to these
singularities in the following section; they are claimed to have
physical consequences in the context of
${\cal PT}$-symmetric Hamiltonians.

\subsection{Brachistochronous behaviour}

The study of quasi-Hermitian operators received a strong boost
when it was realised that Hamiltonians that are symmetric under ${\cal PT}$ 
can have a real spectrum, while not being necessarily Hermitian \cite{bender,benderreview}. 
The question as to physical implications of this more general viewpoint of a formulation of quantum mechanics, 
however, remains open.  

In \cite{brachi1,zg,mostafozcelik} $2\times 2$ quasi-Hermitian matrix Hamiltonians are considered,
in particular the ${\cal PT}$-symmetric Hamiltonian\cite{brachi1} 
\beq
H=\begin{pmatrix} 
re^{i\theta} & s \cr s & re^{-i\theta} 
\end{pmatrix} \; ,
\label{bmat}
\eeq
with $r,s,\theta $ real. It has the eigenvalues 
$E_{\pm}=r \cos \theta \pm \sqrt{s^2-r^2 \sin ^2 \theta}$ and the
unnormalised eigenvectors
\beq
|E_-\rangle =\begin{pmatrix} -\,e^{-i\alpha }\cr 1 
\end{pmatrix} \quad  \quad
|E_+\rangle =\begin{pmatrix} e^{i\alpha }\cr 1 
\end{pmatrix}  \label{eigv}
\eeq
where $\alpha $ is defined by $\sin \alpha =r/s \times \sin \theta $.

This example nicely demonstrates
\begin{enumerate}
\item[(a)] some uniform features of a ${\cal PT}$-symmetric Hamiltonian;
\item[(b)] spontaneous symmetry breaking at an EP,
including all properties inherent to an EP.
\end{enumerate}
\noindent
(a) The spectrum is real when $|s|>|r\sin \theta|$, for this parameter
range the auxiliary parameter $\alpha $ is real, 
$\alpha \in (-\pi/2,\pi/2)$, and
the two eigenfunctions are themselves eigenfunctions of
the ${\cal PT}$ operator. The metric can be chosen as
\beq
\Theta=
\begin{pmatrix} 1 & -i\sin \alpha  \cr i\sin\alpha  & 1
\end{pmatrix} \label{metric}
\eeq
being Hermitian and positive in the range $\alpha \in
(-\pi/2,\pi/2)$. (The general form for $\Theta$ is introduced below, but this choice already
enables us to make some salient points.) 
We mention that the Hamiltonian (\ref{bmat}) when transformed by the
similarity transformation $S=\Theta ^{1/2}$ assumes the form
$$ 
h = SHS^{-1}=\begin{pmatrix} r\cos \theta & 
\sqrt {s^2-r^2\sin ^2 \theta  } \cr
\sqrt {s^2-r^2\sin ^2 \theta  }    &
r\cos \theta
\end{pmatrix} 
$$
which is manifestly Hermitian. Note that this choice of the metric implies that
$\sigma_y$ is quasi-Hermitian, i.e.~an observable, 
but not the $\sigma_z$, as is implied in \cite{brachi1}.

\noindent
(b) When $s=\pm r\sin \theta$, that is when $\alpha =\pm \pi/2$,
the eigenvalues coalesce and we encounter all the properties of an
EP. That is
\begin{enumerate} 
\item[(i)] $H$ cannot be diagonalised, its Jordan form is
$$H=\begin{pmatrix} r\cos \theta & 1 \cr 0  & r\cos \theta 
\end{pmatrix},  $$
\item[(ii)] the eigenvalues are connected by a square root branch point,
\item[(iii)] the eigenvectors become aligned and assume both the form
$$\begin{pmatrix} i \cr 1 \end{pmatrix}$$ and
they cannot be normalised as their norm vanishes,
and 
\item[(iv)] the metric is no longer positive-definite. 
\end{enumerate}
Beyond that point, when
$|s|<|r\sin \theta|$, the eigenvalues are complex, the eigenvectors
are no longer eigenstates of the ${\cal PT}$ operator as $\alpha $ is
complex.

In particular in view of the alignment of the eigenvectors, meaning
physically that there is - at the singularity - just one state, one
cannot speak of a transition. In other words, the concept of a ``transition time''
becomes meaningless.

\subsection{A generalisation}
For the sake of completeness we give here the general form of the metric
obeying (\ref{qh}) for the Hamiltonian (\ref{bmat}). If 
$\sin \alpha <1$ we may choose
\beq
\Theta=
\begin{pmatrix} 1 & \sin \gamma -i\sin \alpha  
\cr \sin \gamma +i\sin\alpha  & 1
\end{pmatrix} \label{gmetric}
\eeq
using the additional free parameter $\gamma $ which must obey the
constraint
\beq \sin ^2\gamma +\sin ^2\alpha <1  \label{ineq} \eeq
to ensure that $\Theta $ is positive-definite. The Hermitian
Hamiltonian generated by $S=\Theta ^{1/2}$ reads now
\beq h = SHS^{-1}=\begin{pmatrix} r\cos \theta & 
f(r,s,\theta ,\gamma ) \cr
\bar f(r,s,\theta ,\gamma )    &
r\cos \theta
\end{pmatrix} \label{gbmat}
\eeq
where
$$f(r,s,\theta ,\gamma )=\frac{s\sin \gamma+ir/s\sin \theta
\sqrt{s^2\cos^2\gamma-r^2\sin^2\theta}}{\sin\gamma+ir/s\sin\theta} $$
and $\bar f$ denotes the complex conjugate.
Of course, it is $f(r,s,\theta ,0)=\sqrt{s^2-r^2\sin ^2\theta }$
and the eigenvalues of (\ref{gbmat}) are still given by those 
of (\ref{bmat}). This generalisation simply confirms that there is
(i) a continuum of choices for a metric, and
(ii) spurious singularities can occur depending on the choice of the
metric. In fact, when $\sin \gamma=\pm \sqrt{1-\sin ^2\alpha }$ the
metric ceases to be positive-definite. When (\ref{ineq}) is
invalidated, $S=\Theta ^{1/2}$ can no longer be Hermitian,
as a consequence (\ref{gbmat}) is not Hermitian either. 
Yet a different choice
for $\gamma $ can remove that deficiency. Note, however, 
even when $\gamma $ is chosen to be zero, the EP
($\sin \alpha =\pm 1$), where
the levels {\it and} the eigenfunctions coalesce, is intrinsic to the problem:
this singularity persists.

Here we mention that, while for $\sin \alpha =1$ the Hamiltonian
(\ref{bmat}) cannot be diagonalised, its hermitised ``counterpart'' is
diagonal and has a genuine degeneracy. However, the two operators are
no longer similarly equivalent since $S^{-1}=\Theta ^{-1/2}$ does not
even exist.

Concerning the question of observables, the general form for the
metric clearly shows that $\sigma_z$ can never be
an observable for (\ref{bmat}). In fact, for this to hold, the
condition $\Theta \sigma _z=\sigma _z \Theta$ had to be obeyed. As
$\Theta $ has the general form $\Theta=1+\sin \gamma \sigma _x+\sin
\alpha \sigma _y$ this can only be for $\alpha =\gamma =0$ reducing
the problem trivially to (\ref{bmat}) being Hermitian. By the same reasoning,
in general, only the combination
$\sin \gamma \,\sigma _x+\sin \alpha \,\sigma _y$ is an observable within
the allowed range of the parameters. Using the appropriate metric $\Theta $,
not only are the expectation values of $\sigma _z$ complex in general
but the probability interpretation becomes inconsistent. In short,
under the regime of (\ref{bmat}) (with $\alpha \ne 0$)
$\sigma_z$ is not an observable. Rather, the $z$-component of spin must be 
described by $\Sigma_z = S^{-1}\sigma_z S$.

\subsection{Classical limit}
We stress that even the classical limit of a non-Hermitian 
${\cal PT}$-symmetric Hamiltonian can depend on the metric chosen \cite{musumbu}. In
fact, the classical limit of (\ref{herm}) is obvious and so is the
classical expression for the energy. The $z$-dependence of the mass
term persists and we obtain $E_{\rm class}=A^2\Omega ^2/(2\mu(z))$
with $A$ being the amplitude of the classical oscillation.

\section{Summary}
\label{sumamry}
Virtually all problems and questions raised in the present paper
become self-evident or obsolete if traditional quantum mechanics is
being adhered to. Yet, recent developments considering quasi-Hermitian
Hamiltonians, in particular in the context of ${\cal PT}$-symmetry,
justify in principle a possible formal extension of the theoretical
framework. However, the consequences (and possible pitfalls) need to
be scrutinised afresh. This is the subject of the present paper.

While there seems to be agreement in the literature that consideration
of quasi-Hermitian Hamiltonians cannot {\it a priori} be excluded on logical or
physical grounds, the non-trivial (non-unity) metric that comes with
it to guarantee a consistent framework is not unique. To be
precise, only when the choice is made of an irreducible set 
of operators $A_i$ satisfying $\Theta A_i^{}=A_i^{\dagger }\Theta $ 
is the metric $\Theta $ fixed.

The examples discussed above illustrate the general points already 
made in Sects. \ref{sectmetric} and \ref{prob}: (i) once a quasi-Hermitian Hamiltonian 
is given, there is some restriction from the start about which traditional observable 
can (or should) be chosen to remain invariant under (\ref{hermobs}), 
and (ii) the set $A_i$ can be a combination of traditional
observables implying one or more continuous parameters. In fact,
$\sigma _z$ cannot be chosen for the Hamiltonian (\ref{bmat}), and
only $(1-z)p^2 +(1+z)x^2$ is a possible choice for (\ref{ham}).
These results have rather stringent consequences. Assuming a physical
setting can be modelled by (\ref{bmat}), the $z$-component of the
spin cannot be described $\sigma_z$; it must in fact be described by
$\Sigma_z = S^{-1}\sigma_z S$. Similarly, the associated orthogonal states describing ``spin up"
and ``spin down", must be the eigenstates of $\Sigma_z$.
Furthermore, certain quantities
of physical interest, such as probabilities and transition amplitudes, will in general
depend on the continuous parameter, that is on $\gamma $ for (\ref{bmat}) and on $z$
for (\ref{ham}). Once the choice (or construction) of an irreducible {\em set} of observables 
has been completed, physical quantities will be fixed, but, as discussed in Sect. \ref{prob} 
some of them will depend on the metric implied by that choice (or construction).

In addition, in contrast to traditionally Hermitian Hamiltonians,
singularities can and generically will occur as part of
the real spectrum. The EPs signal usually the points in parameter
space where (i) the metric becomes singular, and where (ii) the
spectrum ceases to be real. These points of coalescence of spectrum
and eigenfunctions can, because of their specific singular nature,
lead to misinterpretations. The conclusions drawn in the literature
about the Hamiltonian (\ref{bmat}) are doubtful for two reasons.
Firstly, the approach of the
EP, where the eigenspace of (\ref{bmat}) is no longer two-dimensional,
requires particularly cautious analysis. In fact, while it is true that at a
distance from the EP, there is a two-dimensional bi-orthogonal
eigensystem, the normalised vectors tend to infinity and become
aligned when the limit is attained. To translate this into
``transition time going toward zero'' means deviating from quantum
mechanical interpretation where only transitions between strictly
orthogonal and normalisable states can be considered unambiguously\cite{zurek}.
Moreover, since $\sigma _z$ is no valid observable, the spin flip
claimed to be the indicator of a ``brachistochronous'' transition
cannot be observed. This apparently counter-intuitive statement -
after all, can't we always measure a spin projection? - needs
explanation: yes, we can, but {\it not} in terms of $\sigma_z$
when a system should evolve under the
regime of a Hamiltonian as given by (\ref{bmat}). An analysis concerning 
the $z$-component of spin should be carried out in terms of $\Sigma_z = S^{-1}\sigma_z S$
and its eigenstates, in which case the known Hermitian brachistochrone result is again 
obtained\cite{ali2007prl}. 

We feel that the efforts to go beyond the framework of traditional
quantum mechanics has certainly further enhanced our understanding of
it. However, as long as no guidance from some experiments is
at hand, the ambiguities presented here cannot be removed. The
non-uniqueness of the metric and the physical significance of its
singularities require guidance by an actual physical situation. We only
mention that inconsistencies persist when attempts are made
to consider switching from one metric to another, such as by a 
time-dependent metric\cite{ali2007plb}. Such attempts imply changing the underlying Hilbert 
spaces, hence any result can be construed.

\end{document}